\documentclass[twocolumn,preprintnumbers,amsmath,amssymb,pra,superscriptaddress]{revtex4}
\usepackage{amsmath}
\usepackage{amsfonts}
\usepackage{multirow}
\usepackage{float}
\usepackage{graphicx}
\usepackage{nicefrac}
\usepackage{subfigure}
\usepackage{xcolor}
\usepackage{bm}
\usepackage{tikz-cd}
\usepackage[breaklinks=true,pdfborder={0 0 0}]{hyperref}
\usepackage[T1]{fontenc}
\hypersetup{colorlinks=true,citecolor=red}

\begin{document}
\title{Towards superior van der Waals density functionals for molecular crystals}
\author{Dmitry V. Fedorov}
\email{Dm.Fedorov@skoltech.ru}
\affiliation{Skolkovo Institute of Science and Technology, Bolshoy Boulevard 30, 121205 Moscow, Russia}
\author{Nikita E. Rybin}
\affiliation{Skolkovo Institute of Science and Technology, Bolshoy Boulevard 30, 121205 Moscow, Russia}
\affiliation{Digital Materials LLC, Kutuzovskaya street 4A, 143001, Odintsovo, Russia}
\affiliation{National Research Nuclear University MEPhI, Kashirskoe Highway 31, 115409 Moscow, Russia}
\author{Mikhail A. Averyanov}
\affiliation{Skolkovo Institute of Science and Technology, Bolshoy Boulevard 30, 121205 Moscow, Russia}
\affiliation{Mendeleev University of Chemical Technology of Russia, Miusskaya Square 9, 125047 Moscow}
\author{Alexander V. Shapeev}
\affiliation{Skolkovo Institute of Science and Technology, Bolshoy Boulevard 30, 121205 Moscow, Russia}
\affiliation{Digital Materials LLC, Kutuzovskaya street 4A, 143001, Odintsovo, Russia}
\author{Artem R. Oganov}
\affiliation{Skolkovo Institute of Science and Technology, Bolshoy Boulevard 30, 121205 Moscow, Russia}
\author{Carlo Nervi}
\email{C.Nervi@skoltech.ru}
\affiliation{Skolkovo Institute of Science and Technology, Bolshoy Boulevard 30, 121205 Moscow, Russia}
%
\begin{abstract}
Ubiquitous van der Waals (vdW) interactions play a subtle yet crucial role in determining the precise atomic arrangements in
solids, particularly in molecular crystals where these weak forces are the primary link between constituent building blocks.
Within density functional (DF) theory, the most natural approach to address them is the use of vdW-inclusive density
functionals. Through a detailed analysis of the underlying formalism, we have developed a computational scheme that combines
vdW functionals of type DF1 and DF2 providing a readily tunable framework to improve the theoretical description and prediction
of molecular crystals and other sparse materials. The~proof of principle is demonstrated by benchmarking our method on the X23
dataset of molecular crystals.
\end{abstract}
\maketitle

\section{Introduction}
Accurate yet computationally affordable theoretical descriptions of molecular crystals (MCs) are highly desirable in both
established and emerging fields, such as the design of high-energy materials, pharmaceutical compounds, and molecular solids
with tailored properties~\cite{Kole2025}. The structure and stability of these materials are often governed by weak
intermolecular interactions. Although constituting only a small fraction of the total electronic energy, they are decisive in
determining the final crystalline arrangement through their significant forces~\cite{Liu2012,Reilly2015,Kawai2016,Presti2018}.

Among them, the omnipresent van der Waals (vdW) interaction is a subtle quantum-mechanical phenomenon related to electrodynamically
correlated charge fluctuations at different atoms in molecular systems and materials~\cite{Stone-book,Parsegian-book,Kaplan-book}.
Due to their long-range character, it is not so trivial to describe them within the framework of density functional theory (DFT).
Among different approaches to tackle vdW forces within first-principles calculations, the vdW density functionals (vdW-DFs)
attract particular attention~\cite{Lazic2012}, as naturally built within the DFT formalism as well as strongly related
to the robust random-phase approximation (RPA)~\cite{Xinguo2012}. The great progress in the corresponding field of research
has been reviewed in Refs.~\cite{Berland2015,Hyldgaard2020}. In addition, an illuminating interpretation of vdW-DFs was given
by Hyldgaard~\emph{et al.}~\cite{Hyldgaard2014}, who comprehensively discussed the connection between vdW-DFs and the RPA intimately
linked to the many-body dispersion (MBD) method possessing high accuracy~\cite{Tkatchenko2012,DiStasio2014,Ambrosetti2014,Massa2021}. 

In this paper, we provide readers with a concise yet fairly complete description of the related formalism and present an
optimization procedure to design vdW-DFs outperforming existing DFT-based vdW approaches, as demonstrated by comparison
to the lattice energies obtained by the diffusion Monte Carlo (DMC) method~\cite{Pia2024}, for the widely used X23 benchmark
set of MCs~\cite{Otero2012,Reilly2013,Mortazavi2018,Loboda2018,Dolgonos2019}.

\section{Theoretical approach}
The concept of nonlocal vdW-DFs~\cite{First_Comment} has been developed within DFT using a combination of local density
approximation (LDA) and generalized gradient approximation (GGA), by expressing the total exchange-correlation functional
as the following sum~\cite{Berland2015,Hyldgaard2020}
\begin{align}
E_{\rm xc}^{\rm vdW-DF} [n] = E_{\rm xc}^0 [n] + E_c^{\rm nl} [n]\ ,
\label{eq.:E_xc_vdW}
\end{align}
where the semilocal (\emph{outer}) functional
\begin{align}
E_{\rm xc}^0 [n] = E_{\rm x}^{\rm GGA} [n] + E_{\rm c}^{\rm LDA} [n]
= \int d \mathbf{r}\, n (\mathbf{r})\, \mathcal{E}_{\rm xc}^0 (\mathbf{r})
\label{eq.:E_xc_0}
\end{align}
consists of LDA correlation and GGA exchange energy: $\mathcal{E}_{\rm xc}^0 (\mathbf{r}) = \mathcal{E}_{x}^{\rm GGA}
(\mathbf{r}) + \mathcal{E}_{c}^{\rm LDA} (\mathbf{r})$. The explicit separation of nonlocal correlation from its LDA
counterpart is made to avoid double counting of semilocal contributions~\cite{Berland2015,Hyldgaard2020}.

The coupling of plasmon poles corresponding to the electronic structure obtained by means of the outer functional delivers
the nonlocal correlation energy~\cite{Thonhauser2007}
\begin{align}
E_c^{\rm nl} [n] = \frac{1}{2} \iint d \mathbf{r}\, d\mathbf{r}^\prime n(\mathbf{r}) \phi (\mathbf{r},\mathbf{r}^\prime)
n(\mathbf{r}^\prime)\ ,
\label{eq.:E_c_n}
\end{align}
where the kernel
$\phi (\mathbf{r},\mathbf{r}^\prime) = \phi [d (\mathbf{r},\mathbf{r}^\prime), d^\prime (\mathbf{r},\mathbf{r}^\prime)]$
depends on the coordinates through the scaled separations~\cite{Berland2014}
\begin{align}
d (\mathbf{r},\mathbf{r}^\prime) = |\mathbf{r} - \mathbf{r}^\prime| q_{_0} (\mathbf{r})
\ ,\ \ d^\prime (\mathbf{r},\mathbf{r}^\prime) = |\mathbf{r} - \mathbf{r}^\prime| q_{_0} (\mathbf{r}^\prime)\ .
\end{align}
Here, $q_{_0} (\mathbf{r}) = k_{\rm F} (\mathbf{r}) f_x^{\rm in} (s) - (\frac{4\pi}{3}) \mathcal{E}_c^{\rm LDA}
(\mathbf{r})$~\cite{Thonhauser2007} equal to~\cite{Berland2015}
\begin{align}
q_{_0} (\mathbf{r}) = \frac{k_{\rm F} (\mathbf{r})}{\mathcal{E}_x^{\rm LDA} (\mathbf{r})}
\left[ \mathcal{E}_x^{\rm LDA} (\mathbf{r}) f_x^{\rm in} (s) + \mathcal{E}_c^{\rm LDA} (\mathbf{r}) \right]\ ,
\label{eq.:q_0}
\end{align}
with the scaled gradient $s \equiv s (\mathbf{r}) = |\boldsymbol\nabla n (\mathbf{r})|/2 n(\mathbf{r}) k_{\rm F} (\mathbf{r})$
of the electron density $n (\mathbf{r})$ and $k_{\rm F} (\mathbf{r}) = [ 3 \pi^2 n (\mathbf{r}) ]^{\nicefrac{1}{3}}$. In addition,
$\mathcal{E}_x^{\rm LDA} (\mathbf{r}) = - (3/4\pi) k_{\rm F} (\mathbf{r})$ and $\mathcal{E}_c^{\rm LDA} (\mathbf{r})$ is
the (local) correlation of an employed outer functional~\cite{Hyldgaard2020}.

Furthermore, $q_{_0} (\mathbf{r})$ contains the enhancement factor
\begin{align}
f_x^{\rm in} (s) = 1 - (Z_{ab}/9) s^2
\label{eq.:f_x_in}
\end{align}
related to the exchange gradient. Originally, the (\emph{inner}) enhancement factor was taken following an analysis for
screened exchange of Langreth and Vosko~\cite{Langreth1990} who obtained $Z_{ab} = -0.8491$, from a gradient expansion
for the slowly varying electron gas. Their model is assumed to work well for solids with relatively homogeneous electron
distributions. The above value has been employed for vdW-DF1~\cite{Dion2004} but modified later for vdW-DF2~\cite{Lee2010}
to $Z_{ab} = -1.8867$, stemming from a large-$N$ asymptotic analysis~\cite{Schwinger1980,Schwinger1981}, where $N$
is the number of electrons. The modification has been aimed to better describe atoms and molecules~\cite{Elliott2009}.
In addition to $Z_{ab}$\,, distinct vdW-DFs employ different functional forms for their outer functional.

The exchange-correlation energy density in Eq.~\eqref{eq.:q_0},
\begin{align}
\mathcal{E}_{\rm xc}^{\rm in} (\mathbf{r}) = \mathcal{E}_x^{\rm LDA} (\mathbf{r}) f_x^{\rm in} (s) +
\mathcal{E}_c^{\rm LDA} (\mathbf{r})\ ,
\label{eq.:Epsilon_xc_in}
\end{align}
can be also represented in a more general way
\begin{align}
\mathcal{E}_{\rm xc}^{\rm in} (\mathbf{r}) = - \frac{3}{4\pi}\, q_{_0} (\mathbf{r})
= - \frac{q_{_0} (\mathbf{r})}{\pi} \int_{0}^{\infty} \!\! d y\, [1 - h(y)]\ ,
\label{eq.:Epsilon_xc_in_gen}
\end{align}
with $h (y) = 1 - e^{-4\pi y^2/9}$ as the switching function chosen for vdW-DF1/DF2~\cite{Dion2004}, whereas a complex
but smoother and more flexible form of $h (y)$ was employed later for the vdW-DF3 functional~\cite{Chakraborty2020}.
The exchange-correlation energy density is related to the semilocal inner functional
\begin{align}
E_{\rm xc}^{\rm in} [n] = \int d \mathbf{r}\, n (\mathbf{r})\, \mathcal{E}_{\rm xc}^{\rm in} (\mathbf{r})\ ,
\end{align}
which differs from the outer functional by~\cite{Hyldgaard2020}
\begin{align}
\delta E_{\rm x}^0 [n] = E_{\rm xc}^0 [n] - E_{\rm xc}^{\rm in} [n]\ .
\end{align}
The difference reflects a more rapid increase of $\mathcal{E}_{\rm xc}^{\rm in} (\mathbf{r})$ at large values of $s (\mathbf{r})$
in comparison to $\mathcal{E}_{\rm xc}^0 (\mathbf{r})$, which ensures divergence of the scaled separations tuning out
$E_c^{\rm nl}$~\cite{Berland2014}, to avoid spurious contributions from low-density regions~\cite{Hyldgaard2014}.
For each type of vdW-DFs, $\delta E_{\rm x}^0$ is determined by the value of $Z_{ab}$ in $f_x^{\rm in} (s)$ entering
$\mathcal{E}_{\rm xc}^{\rm in} (\mathbf{r})$ of Eq.~\eqref{eq.:Epsilon_xc_in} and by choice of outer exchange-energy density
\begin{align}
\mathcal{E}_{x}^{\rm GGA} (\mathbf{r}) = \mathcal{E}_{x}^{\rm LDA} (\mathbf{r}) F_x (s)\ ,
\label{eq.:E_x_GGA}
\end{align}
defined by the related enhancement factor $F_x (s)$~\cite{Berland2015}.

Thus, various vdW-DFs can be distinguished by the choice of $F_x (s)$ together with the different parametrization of
$f_x^{\rm in} (s)$ encoded in the value of $Z_{ab}$\,. The authors of Ref.~\cite{Berland2015} encouraged the use of
a standard nomenclature vdW-DF-$E_x$ with $E_x$ referring to the exchange functional. Here, we follow their suggestion
but make an extension of the nomenclature to specify also the magnitude of $Z_{ab}$ employed in a scheme: vdW-DF-$E_x$-$Z_{ab}$\,.

To consistently treat exchange energy between outer and inner functionals, in Ref.~\cite{Berland2014} the functional
vdW-DF-cx was introduced, where the enhancement factor
\begin{align}
F_x^{\rm LV-PW86R} (s) = \tfrac{f_x^{\rm in} (s)}{1+\alpha s^6} +
\left(\tfrac{\alpha s^6}{\beta +\alpha s^6}\right) F_x^{\rm PW86R} (s)
\label{eq.:F_x_LV_PW86R}
\end{align}
splines the Langreth-Vosko
gradient expansion~\cite{Langreth1990} of Eq.~\eqref{eq.:f_x_in} with the PW86R~\cite{Second_Comment} enhancement
factor~\cite{Perdew1986,Murray2009}:
$F_x^{\rm PW86R} (s)=(1+a s^2 + b s^4 + c s^6)^{1/15}$, $a=1.851$, $b=17.33$, and $c=0.163$. The spline procedure,
where $\alpha=0.02178$ and $\beta=1.15$~\cite{Berland2014}, ensures \emph{consistent exchange} (cx) treatment
(for small values of $s$). Similar to vdW-DF1 functionals, vdW-DF-cx uses $Z_{ab} = -0.8491$.

In addition, we employ optB88-vdW(-DF1)~\cite{Klimes2010} and rev-vdW-DF2 (also known as vdW-DF-B86R)~\cite{Hamada2014},
both delivering quite accurate results for MCs~\cite{Chakraborty2020}. The outer enhancement factors of the two
(vdW-DF-optB88-0.8491 and vdW-DF-B86R-1.8867 in our nomenclature) functionals are formally the same as the ones
of the related B88 and B86b functionals, differing from their counterparts in two optimized parameters.
Precisely, the vdW-DF-optB88 enhancement factor reads~\cite{Becke1988,Klimes2010}
\begin{align}
F_x^{\rm optB88} (s) = 1 + \mu s^2 / [1 + \beta s\, \text{arcsinh} (cs)]\ ,
\end{align}
where $\mu = 0.22$, $\beta = \mu / 1.2$, and $c = 2 (6 \pi^2)^{1/3}$. This enhancement factor stems from
the original B88 functional of Becke~\cite{Becke1988}, with $\mu = 0.2743$ and $\beta = 9 \mu / 4 \pi$.
Then, the vdW-DF-B86R enhancement factor is given by~\cite{Becke1986,Hamada2014}
\begin{align}
F_x^{\rm B86R} (s) = 1 + \tfrac{\mu s^2}{(1 + \mu s^2 / \kappa)^{4/5}}\ ,
\label{eq.:F_x_B86R}
\end{align}
where $\mu = 10/81$ and $\kappa = 0.7114$. The functional form of Eq.~\eqref{eq.:F_x_B86R} is also valid for its original
counterpart from the B86b functional~\cite{Becke1986}. The corresponding parameters $\mu=0.2449$ and
$\kappa = 0.5757$~\cite{Murray2009} can be obtained from $\beta = 0.00375$ and $\gamma = 0.007$ of Becke~\cite{Becke1986}
through the following relations:
$\mu = \beta(16\pi/3)(6\pi^2)^{\nicefrac{1}{3}} = 4 \gamma \kappa (6\pi^2)^{\nicefrac{2}{3}}$~\cite{Third_Comment}.

We optimize $Z_{ab}$ within the interval bounded by the aforementioned vdW-DF1 and vdW-DF2 counterparts,
as defining the range of reasonable values for this parameter, according to the related limits for the electron
density~\cite{Langreth1990,Dion2004,Lee2010,Schwinger1980,Schwinger1981,Elliott2009}. It is natural to
consider MCs as corresponding to a certain intermediate case with a related in-between value of $Z_{ab}$\,. We also tune
the parameter $\kappa$ in Eq.~\eqref{eq.:F_x_B86R}. The aforementioned original parameters $\beta$ and $\gamma$
of Becke~\cite{Becke1986} were obtained by fitting to noble gas atoms. The values give $\beta / \gamma^{\nicefrac{4}{5}} = 0.20$,
whereas the theoretical large-gradient model gives $\beta / \gamma^{\nicefrac{4}{5}} = 0.27$~\cite{Becke1986}. The latter ratio
delivers $\kappa = 1.0031$, for $F_x^{\rm B86R} (s)$, if the parameter $\mu$ is fixed to its theoretical value of $10/81$.
The obtained parameter set is very close to the one of $F_x^{\rm optB86b} (s) = 1 + \mu s^2/(1+\mu s^2)^{\nicefrac{4}{5}}$,
with $\mu = 10/81$ and $\kappa = 1$, delivering an accuracy similar to $F_x^{\rm optB88} (s)$~\cite{Klimes2011}. We change
$\kappa$ in Eq.~\eqref{eq.:F_x_B86R} between 0.7114 and 1.0031, as a reasonable range for this parameter. In addition,
for all the three considered functionals, the parameter $Z_{ab}$ is tuned in our optimization scheme. Below we show that
the designed functional of the proposed vdW-DF1.5 type delivers accurate results for cell volumes and lattice energies of MCs
from the X23 dataset~\cite{Otero2012,Reilly2013,Mortazavi2018,Loboda2018,Dolgonos2019}. For volumes, we employ the X23b reference
values~\cite{Dolgonos2019}, where phonon contributions (thermal and zero-point energy effects) are \lq\lq removed\rq\rq\ from
experimental data. This comparison ensures the most reasonable benchmark between different (pure electronic) approaches considered
here. Using more accurate reference data (once available in the future) could change a detailed picture, but the proposed
optimization procedure as well as the general arguments for possible improvements of vdW-DFs stay in force. For energies,
we use a stricter comparison procedure employing the corresponding DMC results~\cite{Pia2024}.

\section{Computational details}
Our calculations have been performed by means of the Vienna Ab Initio Simulation Package
(VASP)~\cite{Kresse1993,Kresse1994,Kresse1996_1,Kresse1996_2}, with using setups similar to Refs.~\cite{Loboda2018,Dolgonos2019}
in the case of the X23 set of MCs. The hard projector augmented-wave pseudopotentials were used~\cite{Bloechl1994,Kresse1999}
with an energy cut-off of 1000~eV. The convergence criteria for the geometry optimization correspond to 10$^{-5}$~eV
and 10$^{-3}$~eV$\cdot$\AA$^{-1}$ for the energy and its gradient, respectively. We employed $\Gamma$-centered $k$-point
meshes generated with {\small KSPACING} = 0.3~\AA$^{-1}$, which correspond to the so-called \lq\lq standard\rq\rq\ $k$-point
grids defined in Ref.~\cite{Loboda2018}. The gas-phase calculations were performed with a supercell of
$20\times 20\times 20$~\AA$^3$. The lattice energies were computed by us with the same fixed geometry
(optimized with DFT in Refs.~\cite{Klimes2016,Reilly2013}) as used in Ref.~\cite{Pia2024} to obtain
the corresponding DMC results serving here as the reference. This geometry was also employed for
the benchmark of different DFT-based vdW approaches performed later in Ref.~\cite{Pia2025}.

For the subsequent transferability test, we have considered thirteen ice polymorphs of the DMC-ICE13 data set
(ice Ih, II, III, IV, VI, VII, VIII, IX, XI, XIII, XIV, XV, and XVII)~\cite{Pia2022}. Their lattice energies were
obtained from single-point calculations at the geometries fixed for the corresponding DMC calculations. The crystal
cells contain 6--28 water molecules. An electronic convergence threshold of $10^{-7}$~eV was used. All other computational
settings were the same as for our calculations of the X23 set of MCs. For the considered vdW-DFs, our lattice energies
reproduce the values reported in Ref.~\cite{Pia2022}, with deviations below 1.5~meV per molecule (0.145~kJ/mol).

\begin{figure}[t]
\includegraphics[width=0.85\linewidth]{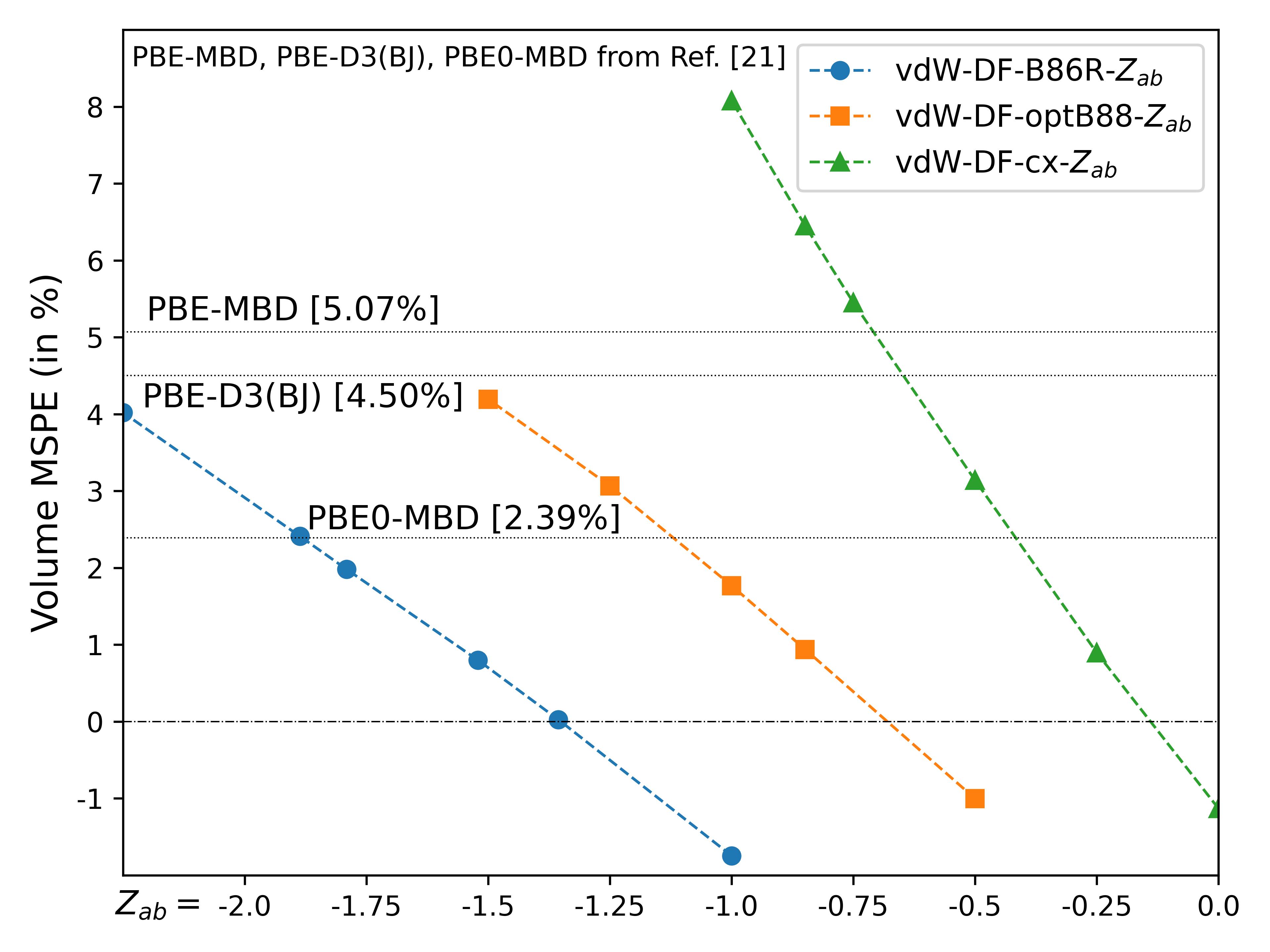}
\caption{Mean signed percentage error for the unit cell volume of the X23 crystals for the considered
vdW-DF-optB88-$Z_{ab}$, vdW-DF-B86R-$Z_{ab}$, and vdW-DF-cx-$Z_{ab}$ functionals. As the reference,
the unit cell volumes $V_{\rm el}^{\rm ref}$ from Ref.~\cite{Dolgonos2019} are taken.}
\label{fig:V_Zab}
\end{figure}

\begin{figure}[b]
\includegraphics[width=0.80\linewidth]{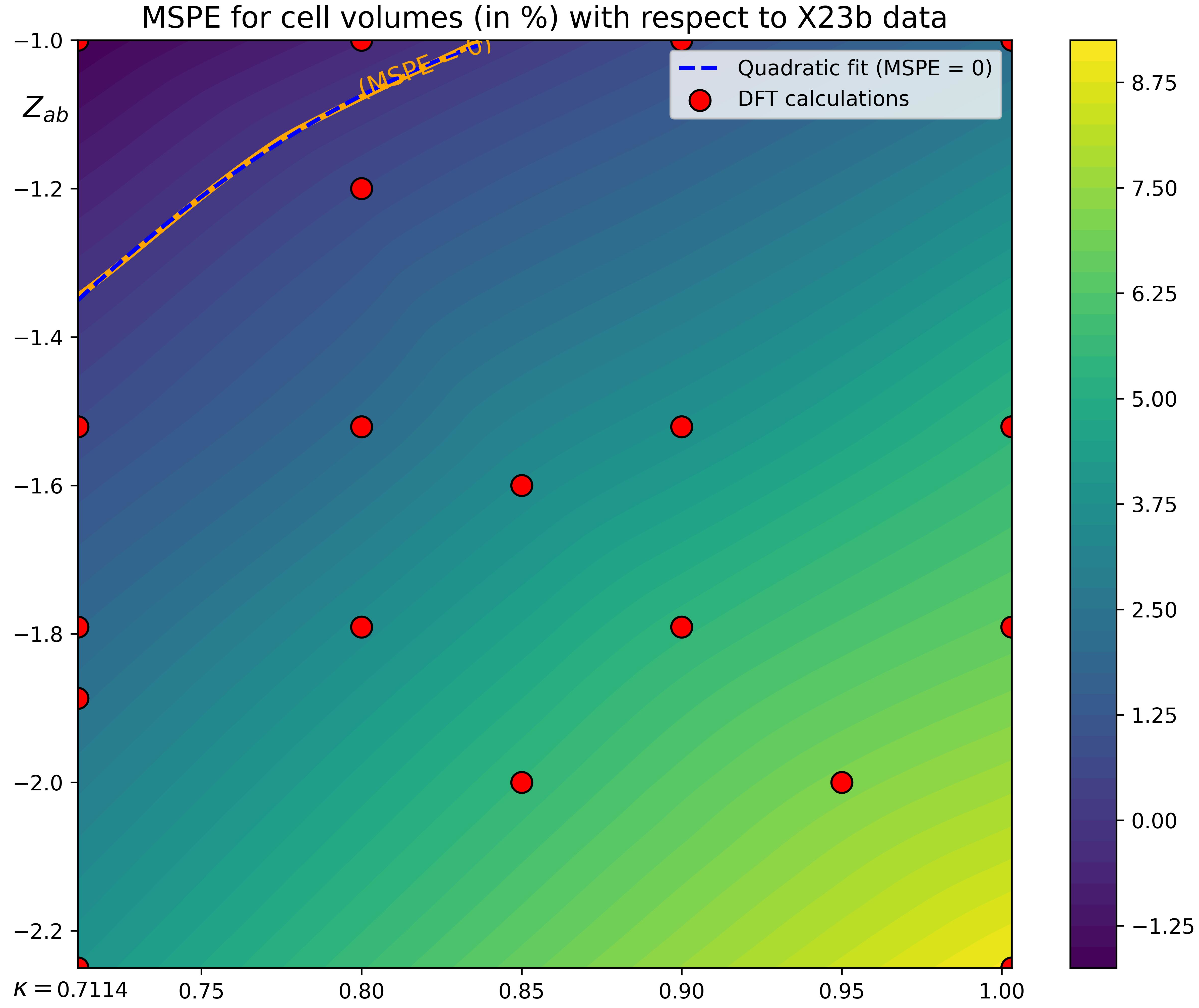}
\caption{MSPE for cell volumes of the X23 set obtained by
the vdW-DF-B86R-$\kappa$-$Z_{ab}$ scheme with respect to the X23b reference data~\cite{Dolgonos2019}. Our DFT results
(red dots) are interpolated to $0.7114 \le \kappa \le 1.0031$ ($x$ axis) and $-2.25 \le Z_{ab} \le -1$~($y$~axis).
For MSPE=0 (the orange curve), the fit of Eq.~\eqref{eq.:Zab_vs_k_V} is shown.}
\label{fig:2D_V_X23b}
\end{figure}

\section{Results}
We first examine the unit cell volumes of MCs from the X23 dataset obtained with the three distinct vdW-DFs by varying
$Z_{ab}$\,. As the reference data, we use the volumes corresponding to the electronic energy surface $V_{\rm el}^{\rm ref}$
derived in Ref.~\cite{Dolgonos2019} by extracting phonon contributions from the experimental data. Thus, we explore
different computational schemes with respect to their quality in describing long-range electronic correlations.
Figure~\ref{fig:V_Zab} shows a monotonic behavior of the mean signed percentage error (MSPE): averaged
$100\% \times (V_{\rm calc} - V_{\rm el}^{\rm ref})/V_{\rm el}^{\rm ref}$\,, where $V_{\rm calc}$ denotes calculated values.
As extended systems, MCs are significantly influenced by the electrodynamic screening~\cite{Schatschneider2013,DiStasio2014},
which can remarkably reduce vdW interactions~\cite{Massa2021}. Increasing the magnitude of $Z_{ab}$ leads to decreasing
the attractive forces and consequently to larger cell volumes calculated with vdW-DFs. The corresponding weaker attraction
in the asymptotic region of vdW-DF2 in comparison to vdW-DF1 was mentioned in Ref.~\cite{Lee2010}. From Fig.~\ref{fig:V_Zab},
the best performance is achieved with vdW-DF-B86R-1.356, vdW-DF-optB88-0.678, and vdW-DF-cx-0.139 delivering the vanishing
MSPE. However, the range of most reasonable values for $Z_{ab}$ is between $Z_{ab} = -0.8491$ and $Z_{ab} = -1.8867$
derived~\cite{Fourth_Comment}, respectively, for solids~\cite{Langreth1990} and atoms/molecules~\cite{Lee2010} as two limits
for condensed matter. With $Z_{ab}=-1.356$ in between of the two above values, the best justified choice corresponds to
vdW-DF-B86R-1.356, for cell volumes. Therefore, below we stay with the vdW-DF-B86R-$Z_{ab}$ scheme of the vdW-DF1.5 type
introduced in this work.

In addition, we change the parameter $\kappa$ in $F_x^{\rm B86R} (s)$ of Eq.~\eqref{eq.:F_x_B86R}. Altogether,
this provides us with the two-dimensional map for the volume MSPE shown in Fig.~\ref{fig:2D_V_X23b}, where the curve
related to the vanishing MSPE can be well approximated by the following quadratic function
\begin{align}
Z_{ab} (\kappa) = -10.043\,\kappa^2 + 18.293\,\kappa - 9.281\ ,
\label{eq.:Zab_vs_k_V}
\end{align}
represented on the map of Fig.~\ref{fig:2D_V_X23b} by the blue dashed line. The corresponding vdW-DF-B86R-$\kappa$-$Z_{ab}$
approach delivers a much more optimizable and efficient computational scheme, as follows from our subsequent discussion.

\begin{figure}[t]
\includegraphics[width=0.80\linewidth]{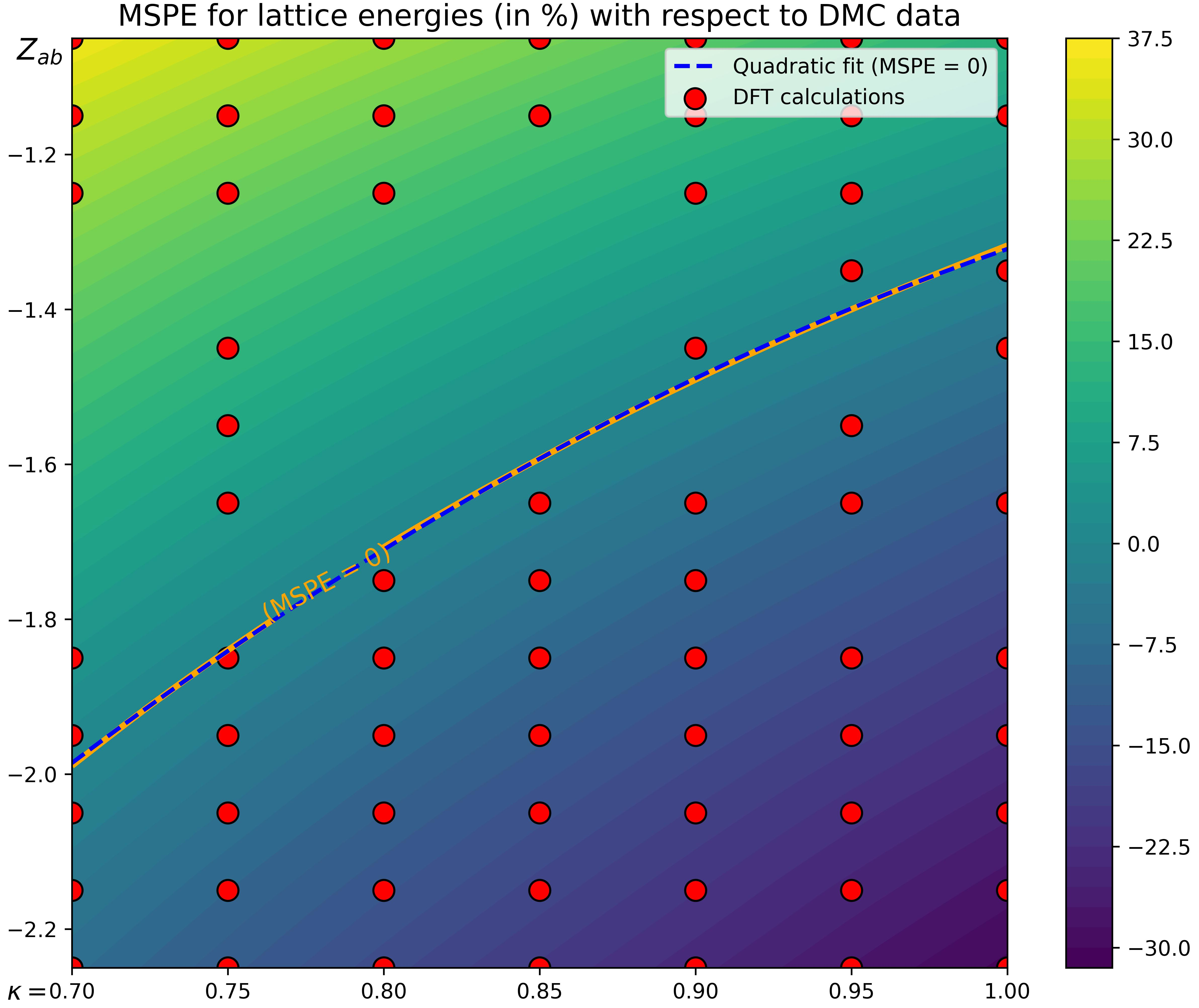}
\caption{MSPE for lattice energies of the X23 crystals obtained by means of
the vdW-DF-B86R-$\kappa$-$Z_{ab}$ scheme with respect to DMC data~\cite{Pia2024}.
Our DFT results (red dots) are interpolated to $0.70 \le \kappa \le 1.00$ ($x$ axis) and
$-2.25 \le Z_{ab} \le -1.05$~($y$~axis). For MSPE=0 (the orange curve), the fit of Eq.~\eqref{eq.:Zab_vs_kappa} is shown.}
\label{fig:Inter_DMC}
\end{figure}

\begin{figure}[b]
\includegraphics[width=0.95\linewidth]{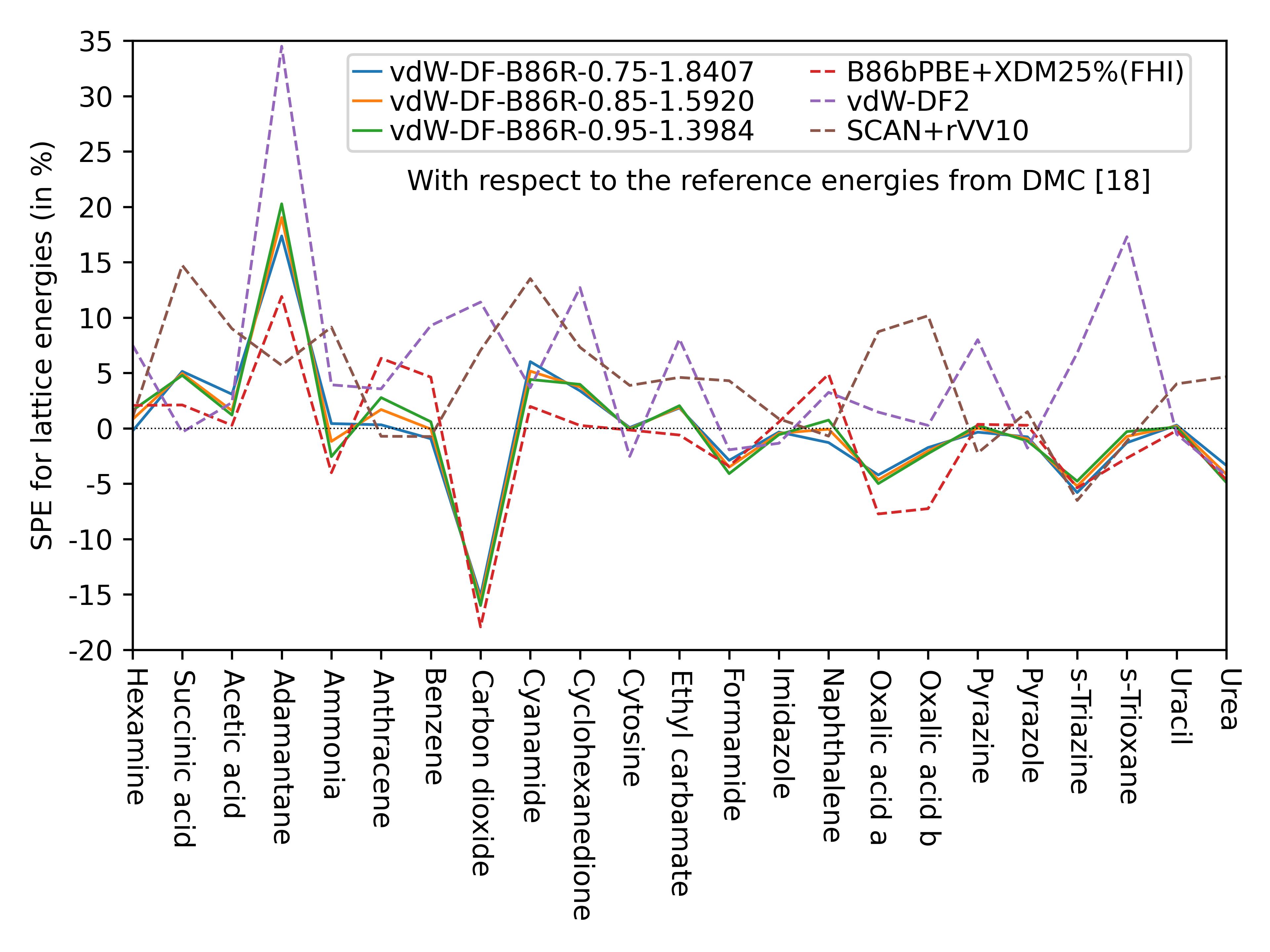}
\caption{Signed percentage error for lattice energies of X23 molecular crystals obtained by our
vdW-DF-B86R-$\kappa$-$Z_{ab}$ scheme with three different pairs of parameters $\{\kappa, Z_{ab}\}$ compared
to other efficient approaches benchmarked in Ref.~\cite{Pia2025}.}
\label{fig:E_seven}
\end{figure}

For our analysis of lattice energies of the X23 MCs, we use the related DMC data~\cite{Pia2024} as a more
robust reference, with pure electronic contributions unaffected by phonons. Figure~\ref{fig:Inter_DMC} shows
a two-dimensional map (similar to Fig.~\ref{fig:2D_V_X23b}), where the curve related to the vanishing MSPE
can be well approximated by the following quadratic function
\begin{align}
Z_{ab} (\kappa) = -2.7560\,\kappa^2 + 6.8966\,\kappa - 5.4629\ ,
\label{eq.:Zab_vs_kappa}
\end{align}
represented on the map by the (blue) dashed line. Here, the conventional value $\kappa = 0.7114$ leads to $Z_{ab} = - 1.9514$
which lies outside the aforementioned range of its reasonable values. Our switch from the vdW-DF-B86R-$Z_{ab}$ to
vdW-DF-B86R-$\kappa$-$Z_{ab}$ scheme (by changing $\kappa$ in addition to $Z_{ab}$) becomes really desirable and useful.
Keeping the reasonable range of $\kappa$ values ($0.7114$ to $1.0031$) and following Eq.~\eqref{eq.:Zab_vs_kappa},
we consider the pairs $\{0.75, -1.8407\}$, $\{0.85, -1.5920\}$, and $\{0.95, -1.3984\}$, for $\{\kappa, Z_{ab}\}$ set.
The corresponding results presented in Fig.~\ref{fig:E_seven} and Table~\ref{tab.:Table_LE} show similar accuracy
between the three schemes, which are also compared to some best approaches of Ref.~\cite{Pia2025}.

Namely, Fig.~\ref{fig:E_seven} presents the signed percentage error (SPE) for the lattice energy of each MC from the X23 set.
Clearly, the proposed vdW-DF-B86R-$\kappa$-$Z_{ab}$ schemes work better than vdW-DF2 and SCAN+rVV10 methods showing much
larger SPE amplitudes in most cases. On average, our results are visually comparable to the ones obtained by B86bPBE+XDM25\%(FHI)
identified as the best approach within the benchmark of Ref.~\cite{Pia2025}. However, the detailed statistical analysis of
Table~\ref{tab.:Table_LE} shows that our vdW-DF-B86R-$\kappa$-$Z_{ab}$ approach even outperforms it. Among our three schemes,
a slightly better performance is delivered by vdW-DF-B86R-$0.75$-$1.8407$, which we recommend to use as the best functional
of the proposed vdW-DF1.5 type, for the X23 MCs. This functional remarkably outperforms the vdW-DF2 functional shown to be
the best vdW-DF approach with respect to the reference DMC results~\cite{Pia2024}, as was identified within the benchmark
performed on the X23 set of MCs in Ref.~\cite{Pia2025}.

\begin{table*}
\centering
\begin{ruledtabular}
\begin{tabular}{c c c c c c}
Our results  &    MAE~(kJ/mol)  &  ME~(kJ/mol)  &  RMSE~(kJ/mol)  &  MAPE~(\%)  &  MSPE~(\%) \\
\hline
vdW-DF-B86R-$0.75$-$1.8407$~~~~~~~~~~~~~~~~~~~~~~~    &  ~2.32  &   ~0.27  &  ~3.43  &  ~3.32  &  $-$0.02 \\
vdW-DF-B86R-$0.85$-$1.5920$~~~~~~~~~~~~~~~~~~~~~~~    &  ~2.39  &   ~0.29  &  ~3.60  &  ~3.38  &   ~0.02  \\
vdW-DF-B86R-$0.95$-$1.3984$~~~~~~~~~~~~~~~~~~~~~~~    &  ~2.60  &   ~0.31  &  ~3.80  &  ~3.67  &   ~0.05  \\
vdW-DF-B86R-$0.7114$-$1.356$~~~~~~~~~~~~~~~~~~~~~    &  16.85  &   16.85  &  18.32  &  20.96  &   20.96  \\
vdW-DF-B86R-$0.7114$-$1.8867$ [rev-vdW-DF2]           &  ~2.84  &   ~1.86  &  ~4.02  &  ~3.85  &   ~1.98  \\
\hline
Best results of Ref.~\cite{Pia2025}        &           MAE~(kJ/mol)  &  ME~(kJ/mol)  &  RMSE~(kJ/mol)  &  MAPE~(\%)  &  MSPE~(\%) \\
\hline
B86bPBE+XDM25\%(FHI)          &    2.73  &   $-$0.39 &   3.73  &  3.90 &  $-$0.80  \\
B86bPBE+XDM(FHI)              &    2.89  &   ~0.63   &   3.87  &  4.10 &  ~0.73    \\
B86bPBE+XDM50\%(FHI)          &    3.10  &   $-$1.07 &   3.73  &  4.50 &  $-$1.96  \\
B86bPBE+XDM(QE)               &    3.25  &   ~0.30   &   4.32  &  4.69 &  ~0.39    \\
vdW-DF2                       &    4.40  &   ~3.19   &   6.34  &  6.38 &  ~5.27    \\
SCAN+rVV10                    &    4.58  &   ~3.91   &   6.28  &  5.31 &  ~4.26    \\
PBE+MBD                       &    4.70  &   ~2.46   &   5.81  &  6.60 &  ~2.76    \\
PBE0+MBD(FHI)                 &    5.72  &   ~4.25   &   6.99  &  7.50 &  ~4.73    \\
\end{tabular}
\end{ruledtabular}
\caption{Benchmark of DFT methods with respect to the reference DMC lattice energies, E[DMC]~\cite{Pia2024}:
MAE = Mean($|\Delta|$), ME = Mean($\Delta$), RMSE = $\sqrt{\text{Mean}(\Delta^2)}$, MAPE = Mean($|\Delta|$/E[DMC])$\times$100\%,
and MSPE = Mean($\Delta$/E[DMC])$\times$100\%\,, where the term \lq\lq Mean\rq\rq\ denotes an averaging over all the molecular
crystals within the X23 dataset and $\Delta$ = $E$[Method] - $E$[DMC].}
\label{tab.:Table_LE}
\end{table*}

In addition, we performed the geometry relaxation of the X23 MCs by the three best approaches of Table~\ref{tab.:Table_LE}.
With respect to the reference volumes from Ref.~\cite{Dolgonos2019}, vdW-DF-B86R-$0.75$-$1.8407$, vdW-DF-B86R-$0.85$-$1.5920$,
and vdW-DF-B86R-$0.95$-$1.3984$ deliver the MSPE of 3.11\%, 3.54\%, and 3.96\%, respectively. The above three schemes show
better performance than the widely used PBE-D3(BJ) [4.50\%] or PBE-MBD [5.07\%]~\cite{Loboda2018}. Moreover, the average
volumetric expansion due to the zero-point effects (at 0\,K) obtained in Ref.~\cite{Dolgonos2019} by means of PBE-D3(BJ)
is 2.4\%, for the X23 MCs. Thus, the presented functionals serve as an efficient approach simultaneously providing
highly accurate lattice energies as well as unit cell volumes (at zero temperature, including the volumetric expansion
due to the zero-point effects).

By comparing Figs.~\ref{fig:2D_V_X23b} and \ref{fig:Inter_DMC}, one can draw several observations relevant
to further applications of our proposed approach. The magnitude of the energy MSPE is much more
sensitive to the change of the $\{\kappa, Z_{ab}\}$ parameters than its volume counterpart. Therefore, in practice
it is better to optimize vdW-DF-B86R-$\kappa$-$Z_{ab}$ schemes first with respect to lattice energies and then
to volumes (by an additional fine tuning). An opposite way causes large errors in the lattice energy, as it is
demonstrated by the results of Tables~\ref{tab.:Table_LE} and \ref{tab:ice13} obtained with the vdW-DF-B86R-$0.7114$-$1.356$
scheme. In our variation of the two parameters, the cell volumes (on average) become overestimated/underestimated in
the opposite way to the lattice energies. Finally, the corresponding curves for the vanishing MSPE are practically
\emph{parallel} to each other. Altogether, this suggests the possibility of developing a vdW-DF that reduces the bias
in both properties, thereby making the two curves defined by the quadratic functions in Eqs.~\eqref{eq.:Zab_vs_k_V} and
\eqref{eq.:Zab_vs_kappa} coincide. In such a case, changing the parameters along just a single (MSPE = 0) curve would
provide an additional degree of freedom to perharps better describe other system properties. This idea points to a promising
direction for further research.

Our findings that, for both the unit cell volumes and the lattice energies of the MCs from the X23 set, the performance of
the considered functional is improved by choosing $Z_{ab}$ between its DF1 and DF2 values evidently support the vdW-DF1.5
formalism proposed in this paper. The optimization procedure, which delivers the new set of corresponding vdW-DF-B86R-$\kappa$-$Z_{ab}$
functionals of vdW-DF1.5 type can be applied in a similar way to any other kind of sparse materials. Like for the considered
case of X23 MCs, it might require searching for different optimized parameters $\{\kappa, Z_{ab}\}$ to properly describe
both, the geometry and energy properties. Such a fine tuning is assumed to provide one with the best performance of the
proposed approach. However, one can foresee a future development of a universal vdW-DF as a superposition of many distinct
vdW-DF-B86R-$\kappa$-$Z_{ab}$ combined within a single neural network scheme, that is equivalent to finding a universal
analytical functional for DFT formalism.

To demonstrate the transferability of the optimized functionals beyond the X23 set used for their construction, we also
compute the lattice energy of the thirteen ice polymorphs of the DMC-ICE13 data set (ice Ih, II--IV, VI--IX, XI, XIII--XV,
and XVII)~\cite{Pia2022}. Ice serves as a good test system because its cohesion is dominated by a hydrogen-bonded network.
Its polymorphs have different densities: from the open forms ice Ih and XVII ($\sim$30~\AA$^3$ per molecule) to
the compact high-pressure phases VII and VIII ($\sim$20~\AA$^3$ per molecule). Our calculations were performed at
the fixed PBE geometries of the DMC reference found to deliver good agreement with experimental structures~\cite{Pia2022}.
Table~\ref{tab:ice13} summarizes the obtained results.

\begin{table*}
\centering
\begin{ruledtabular}
\begin{tabular}{l c c c c c}
Our calculations & MAE~(kJ/mol) & ME~(kJ/mol) & RMSE~(kJ/mol) & MAPE~(\%) & MSPE~(\%) \\
\hline
vdW-DF-B86R-$0.75$-$1.8407$   &   4.61  &  4.61  &  4.79  &  7.98  &   7.98 \\
vdW-DF-B86R-$0.85$-$1.5920$   &   3.61  &  3.61  &  3.80  &  6.24  &   6.24 \\
vdW-DF-B86R-$0.95$-$1.3984$   &   2.76  &  2.76  &  2.98  &  4.77  &   4.77 \\
vdW-DF-B86R-$0.7114$-$1.356$  &  10.88  & 10.88  & 10.90  & 18.89  &  18.89 \\
\hline
Results of Ref.~\cite{Pia2022}  &  MAE~(kJ/mol)  &  ME~(kJ/mol)  &  RMSE~(kJ/mol)  &  MAPE~(\%)  &  MSPE~(\%) \\
\hline
optB88-vdW                                   &   9.33  &   9.33  &  9.35  & 16.22  &  16.22   \\
optB86b-vdW [vdW-DF-B86R-$1.0$-$0.8491$]     &   7.05  &   7.05  &  7.08  & 12.26  &  12.26   \\
optPBE-vdW                                   &   9.20  &   9.20  &  9.22  & 16.01  &  16.01   \\
rev-vdW-DF2 [vdW-DF-B86R-$0.7114$-$1.8867$]  &   5.71  &   5.71  &  5.85  &  9.89  &   9.89   \\
vdW-DF2                                      &   1.34  &   1.34  &  1.62  &  2.36  &   2.36   \\
vdW-DF                                       &   4.91  & $-$4.91 &  4.97  &  8.52  &  $-$8.52 \\
\end{tabular}
\end{ruledtabular}
\caption{Benchmark of vdW DFs with respect to the reference DMC lattice energies, E[DMC]~\cite{Pia2022},
for the 13 DMC-ICE13 ice polymorphs: MAE = Mean($|\Delta|$), ME = Mean($\Delta$), RMSE = $\sqrt{\text{Mean}(\Delta^2)}$,
MAPE = Mean($|\Delta|/E[\mathrm{DMC}]$)$\times$100\%, and MSPE = Mean($\Delta/E[\mathrm{DMC}]$)$\times$100\%, where
$\Delta = E[\mathrm{Method}] - E[\mathrm{DMC}]$ and the mean is over all 13 ice polymorphs.}
\label{tab:ice13}
\end{table*}

The best three vdW-DF-B86R-$\kappa$-$Z_{ab}$ schemes of Table~\ref{tab.:Table_LE} (coming from our X23 optimization) overbind
the polymorphs, with the mean absolute errors (MAEs) of 4.61~kJ/mol, 3.61~kJ/mol, and 2.76~kJ/mol for vdW-DF-B86R-$0.75$-$1.8407$,
vdW-DF-B86R-$0.85$-$1.5920$, and vdW-DF-B86R-$0.95$-$1.3984$, respectively. Their MAEs are less than those of the vast
majority of other vdW-DFs considered in Ref.~\cite{Pia2022} such as rev-vdW-DF2 (5.71~kJ/mol), optB88-vdW (9.33~kJ/mol)
and optB86b-vdW (7.05~kJ/mol), optPBE-vdW (9.20~kJ/mol), and vdW-DF (4.91~kJ/mol). The only vdW-DF which is more accurate
than our best three vdW-DF-B86R-$\kappa$-$Z_{ab}$ schemes is the vdW-DF2 functional (with the MAE of 1.34~kJ/mol). However,
this functional overbinds the lattice energy of X23 crystals with MAE of 4.40~kJ/mol (see Table~\ref{tab.:Table_LE}),
whereas our vdW-DF-B86R-$0.95$-$1.3984$ scheme keeps both MAEs below 3~kJ/mol (namely, 2.60~kJ/mol for X23 and 2.76~kJ/mol for
ice). For the used benchmark sets, our functional also outperforms the PBE0+MBD approach (with 5.72~kJ/mol and 4.12~kJ/mol for
X23~\cite{Pia2025} and ice~\cite{Pia2022}, respectively). Moreover, the difference in the accuracy of the aforementioned
approaches becomes even more remarkable, if one considers MAPE/MSPE for the X23 MCs: vdW-DF-B86R-$0.95$-$1.3984$ with
(3.67/0.05)\% versus the vdW-DF2 functional with (6.38/5.27)\% or PBE0+MBD with (7.50/4.73)\%. Thus, one can conclude
that the transferability of our approach is justified. Here, it is worth emphasizing that within our current study
the vdW-DF-B86R functional form kept unchanged and the corresponding parameters are optimized only. Just such a simple
optimization already delivers a highly accurate and transferable approach. In its turn, an additional proper modification
of the functional form (to be found in the future) can be foreseen as a great opportunity for further improvements in
accuracy and transferability.

During our preparation of this manuscript, another approach to improve vdW-DFs for MCs was published by
Jenkins~\emph{et al.}~\cite{Jenkins2025}.~The proposed vdW-DF3-mc functional has fixed $Z_{ab} = -1.887$
but increased flexibility of nonlocal correlation through reparametrization of the complex switching
function of the corresponding vdW-DF3 type~\cite{Chakraborty2020} via revoking the constraint related to
Eq.~\eqref{eq.:Epsilon_xc_in_gen}. The optimized DF3 parametrization was obtained by comparison of results
obtained by means of the {\small QUANTUM ESPRESSO} package~\cite{Giannozzi2017} with experimental data, to
describe MCs at finite temperature. By contrast, our vdW-DF1.5 approach is developed to capture the crystal
properties at 0\,K (either neglecting the phonon contributions at all or including just the volumetric
expansion due to the zero-point effects). In addition, we keep the switching function of the vdW-DF1/DF2
type with the constraint of Eq.~\eqref{eq.:Epsilon_xc_in_gen} fulfilled. Thus, our work and the one of
Ref.~\cite{Jenkins2025} serve as two complementary studies, and their combination can eventually lead
to a successful development of highly accurate vdW-DFs designed for MCs, with a possible extension
to any kind of other sparse materials.

We share the view expressed in Ref.~\cite{Berland2015} on the probable future replacement of dispersion-corrected
DFT by nonlocal vdW-DFs. Indeed, based on the results of Jenkins~\emph{et al.}~\cite{Jenkins2025} as well as our
work presented here, further intensive studies in this direction are anticipated, in order to develop advanced
vdW-DFs capable of describing sparse materials, where vdW forces can significantly act through important low-density
regions~\cite{Hyldgaard2020}. The relevant materials are not restricted to molecular crystals but belong to
vast classes of systems such as layered materials, polymers, and covalent or metal-organic frameworks.

In summary, we introduce vdW-DF1.5 as a new type of vdW-DFs, which is a combination of the DF1 and DF2 types with respect
to the screening parameter $Z_{ab}$ in the enhancement factor of the inner functional. Its changing together with possible
tuning parameters of the outer enhancement factor (for GGA exchange) results in the powerful optimization procedure
potentially applicable to any kind of sparse materials. The proof of principle is demonstrated on the X23 set of molecular
crystals, for which our optimized vdW-DF-B86R-$\kappa$-$Z_{ab}$ scheme outperforms other efficient DFT-based approaches.
The transferability of the developed scheme is illustrated on the thirteen ice polymorphs. The presented results provide
the community with a number of observations helpful for the development of vdW-DFs as well as with the practical procedure
to improve already existing functionals, for a natural and efficient way of tackling the subtle long-range correlations
within the DFT framework.

The work was supported by the grant for research centers in the field of AI provided by the Ministry of Economic
Development of the Russian Federation in accordance with the agreement 000000C313925P4F0002 and the agreement with
Skoltech \textnumero139-10-2025-033 as well as by the Russian Science Foundation within the project 25-43-20043.

\bibliographystyle{apsrev4-1}
\bibliography{literature}

\end{document}